\documentclass[12pt,fleqn]{article}
\usepackage{amsfonts,lineno}
\usepackage{graphicx}
\usepackage{amsmath,amsthm,amssymb,amscd,multirow}
\usepackage{array,longtable}
   \numberwithin{equation}{section} \numberwithin{theorem}{section} \numberwithin{lemma}{section}
 \numberwithin{corollary}{section} \allowdisplaybreaks

\newcommand{\beq}{\begin{equation}}
\newcommand{\eeq}{\end{equation}} \newcommand{\bsp}{\begin{split}} \newcommand{\esp}{\end{split}}

\begin{document}
\bibliographystyle{Vancouver}

 \title{Bayesian Detection of Abnormal ADS in Mutant \textit{Caenorhabditis elegans} Embryos}
 \author{Wei Liang$^a$, Yuxiao Yang$^a$, Yusi Fang$^a$, Zhongying Zhao$^b$, Jie Hu$^{a*}$} 
 
 
 
 \date{\begin{flushleft}
 		\begin{small}
 			a:School of Mathematical Sciences, Xiamen University, Xiamen, China\\
 			b:Department of Biology, Hong Kong Baptist University, Hong Kong, China\\
 			*:hujiechelsea@xmu.edu.cn
 		\end{small} 		
\end{flushleft}}
\maketitle \maketitle

\begin{abstract}
Cell division timing is critical for cell fate specification and morphogenesis during embryogenesis. How division timings are regulated among cells during development is poorly understood. Here we focus on the comparison of asynchrony of division between sister cells (ADS) between wild-type and mutant individuals of \textit{Caenorhabditis elegans}. Since the replicate number of mutant individuals of each mutated gene, usually one, is far smaller than that of wild-type, direct comparison of two distributions of ADS between wild-type and mutant type, such as Kolmogorov- Smirnov  test, is not feasible. On the other hand, we find that sometimes ADS is correlated with the life span of corresponding mother cell in wild-type. Hence, we apply a semiparametric Bayesian quantile regression method to estimate the $95\%$ confidence interval curve of ADS with respect to life span of mother cell of wild-type individuals. Then, mutant-type ADSs outside the corresponding confidence interval are selected out as abnormal one with a significance level of $0.05$. Simulation study demonstrates the accuracy of our method and Gene Enrichment Analysis validates the results of real data sets.

\textbf{Keywords:} ADS, Quantile regression, MCMC

\end{abstract}

\section{Introduction}
How a single-celled zygote develops into a mature embryo with different tissues is still a foundational but pendent conundrum in developmental biology. Without question, gene expression pattern and dynamics play a crucial role in this procedure.

The embryogenesis of \textit{C. elegans} has been studied extensively in the past several decades, especially after development of high throughout automated experimental techniques were intruduced. \cite{Bao2006} and \cite{Murray2008} developed an automated system to analyze the continuous gene expression profiles in \textit{C. elegans} with cellular resolution from zygote till mid-embryogenesis using time-lapse confocal laser microscope. With this system, \cite{Murray2012,Long2009,Liu2009,Spencer2011,Ho2015,Hu2015} analyzed the expression patterns of various genes and their relationships with cell fates and tissue differentiation. Among them, \cite{Ho2015} stated that there're reasons to believe that ADS can result in cell-specific division pattern and hence affect tissue formation. Previous study of cell division timing mainly focused on heterochronic genes during postembryonic stage of \textit{C. elegans}, such as \cite{Ren2010,Gleason2010}. However, the mechanisms of cell division in embryonic and postembryonic stages are quite different \cite{Ambros2001}. With the help of newly collected data sets by \cite{Bao2006,Murray2008}, we can analyze the cell division timing systematically and quantitatively. 

Here, we aim to distinguish whether the ADS of mutant type is abnormal. Since various mother cells demonstrate various types of cell division and ADS, we use mother cells to label different types of ADS in the following content. Due to 260 individuals of wild-type but only one copy of each mutation type, direct comparison of distributions between wild-type and target mutant type is not feasible. On the other hand, the above-mentioned confocal microscopy on \textit{C. elegans} embryogenesis is designed to measure the expression level of one specific target gene on all existing cells of an individual embryo during development. Due to strain differences (such as the insertion of DNA sequence coding fluorescent protein into various locations of the \textit{C. elegans} genome) and variability in experimental and environmental factors, even ADS of the same mother cell in different wild type individuals show high quantitative variation, indicating considerable noise. But the lack of principled statistical methods hinders the comprehensive understanding of these data sets. For example, \cite{Ho2015} ignored the relationships between ADS and life time of corresponding mother cell, and used an ad hoc threshold to report the abnormal ADS. 

In order to overcome these difficulties, we apply a Bayesian semiparametric quantile regression method to classify abnormal ADS in mutants. Bayesian quantile regression, which combines the advantages of quantile regression and Bayesian method, has been studied over a period of time and applied wildly in research areas, such as \cite{Yu2001,Dunson2005,Kottas2009,Thompson2010}. Our method is based on kernel function suggested by \cite{Takeuchi2005} to estimate the $95\%$ confidence interval curve of wild-type ADS for each mother cell. For more robust and stable estimation, we transform the optimization problem into a Bayesian framework with MCMC algorithm applied to obtain the estimation. Then we can easily classify the mutant-type ADS based on whether it is outside the corresponding confidence interval.

In Section~\ref{method}, we present our Bayesian semiparametric quantile regression algorithm and the hypothesis testing framework of whether mutant-type ADS is abnormal compared to wild-type. In Section~\ref{application}, we synthesize some mimic data sets to validate our algorithm and apply the algorithm to real data files. Section~\ref{conclusion} concludes the paper.

\section{Methodology}\label{method}

\subsection{Experimental Data}\label{data}
\begin{figure}[htbp]
\centering
\includegraphics[scale=0.5]{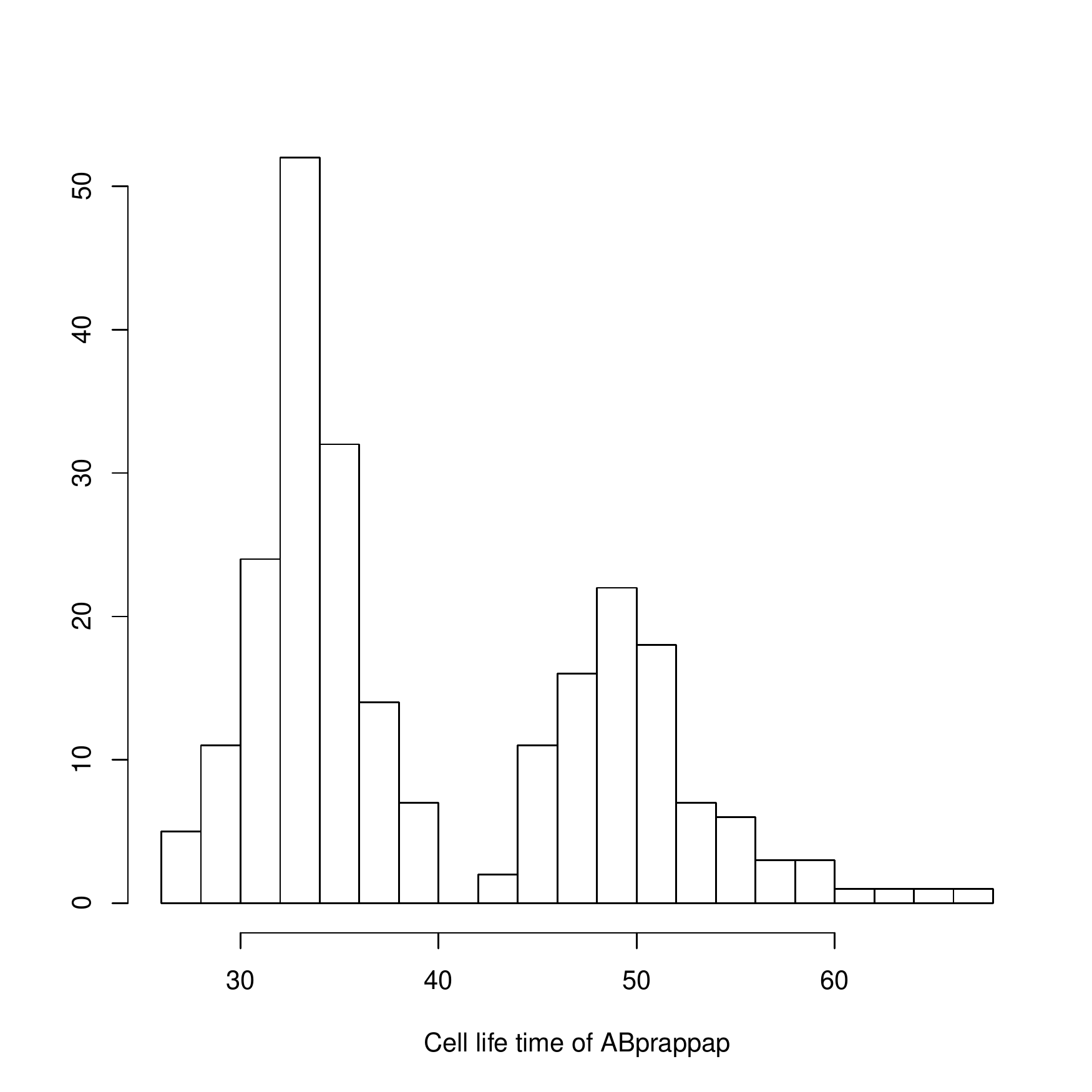}
\caption{Histogram of life time of cell ABprappap.}\label{lifetime}
\end{figure}

Our real data application is based on the data provided by \cite{Murray2012,Ho2015}. It consists of 260 wild-type individuals and 83 mutant-type ones. Division timing of each cell is measured from its birth to its death with teporal resolution of approximately 1.5 minutes. So all the life times of cells were measured as discrete positive integer due to the technical restriction. For example, a measurement of $9$ minutes of life time may be $7.5$ to $10.5$ minutes in reality. But it doesn't matter to our study since all the times no matter wild-type or mutant type, are measured in the same way. It's also worth noting that the distribution of life time of a given cell among wild-type can be multimodal and skewed as Figure~\ref{lifetime} shown.

\subsection{Assumptions}\label{assumption}

In total, we have 351 different mother cells corresponding to various types of ADS among wild-type individuals. We denote these data sets of ADS and corresponding life time of mother cell as $\mathcal{D}_1,\,\mathcal{D}_2,\,\cdots,\,\mathcal{D}_{351}$. Each of them consists of about 260 data points collected from our 260 wild-type individuals, and each data point has two values, ADS, denoted by $X$, and life time of corresponding mother cell, denoted by $Y$. For every data $\mathcal{D}_i$, we run Pearson's test and Spearman's test (see Appendix~\ref{a1} for details) to see whether there're linear or other monotonic relationship between $X$ and $Y$. Table~\ref{test} shows the testing results with $\alpha=0.05$.
\begin{table}
\begin{tabular}{c|c|c}
\hline
Test&No. of Failure &Proportion of Failure\\
\hline
Pearson's Test&99&38.98$\%$\\
\hline
Spearman's Test&119&46.85$\%$\\
\hline
both Pearson's Test and Spearman's Test&90&35.43$\%$\\
\hline
\end{tabular}
\caption{Results of Pearson's Test and Spearman's Test where $H_0$ means no given relationship between $X$ and $Y$.}\label{test}
\end{table}

\begin{figure}[htbp]
\centering
\includegraphics[scale=0.45]{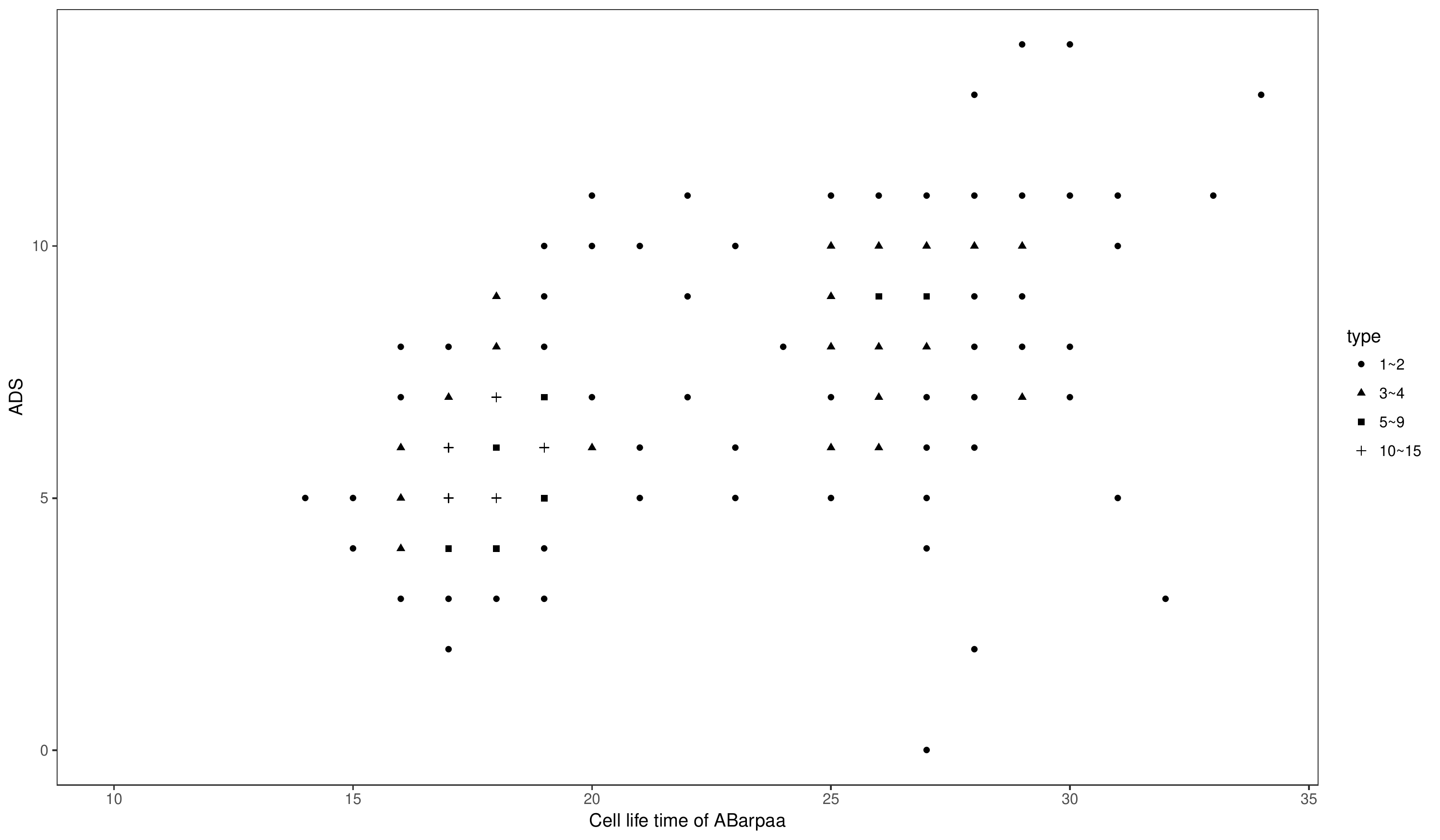}
\caption{An example of data set $\mathcal{D}_i$. X-axis represents life time of mother cell ABarpaa and Y-axis represents ADS. Different shapes of points indicate different frequency as labeled.}\label{linear}
\end{figure}

We can see that in nearly half of $\mathcal{D}_i$, we should reject null hypothesis and there're probably some monotonic relationship between $X$ and $Y$. On the other hand, 90 data sets of $\mathcal{D}_i$ reject $H_0$ in both tests which means most of the monotonic relationships may be linear. Figure~\ref{linear} is an example of data set $\mathcal{D}_i$ which reject $H_0$ in both tests.

Based on this analysis, we assume a linear function as parametric part in our quantile regression model and details are shown in Section~\ref{model}.

\subsection{The Bayesian Modelling Framework for Quantile Regression}\label{model}

In this section, we introduce a semiparametric quantile regression within Bayesian framework for the ADS data of wild-type individuals.

Since we assume that $X$ and $Y$ may be monotonically related, we model the quantile function $f_{\tau}(x)$ as the following semiparametric form,
\begin{equation*}
f_{\tau}(x)=w_{\tau}(x)+\beta_{\tau} x,
\end{equation*}
where $\beta_{\tau}x$ is the parametric component, $w_{\tau}(x)$ is the nonparametric component. 

Next, we choose the following Gaussian kernel to model the unknown function $w_{\tau}(x)$, that is
\begin{equation*}
w_{\tau}(x)=\sum_{i=1}^n\alpha_jk(X_j,\,x)+c,
\end{equation*}
where $k(x,\,x')=\exp{(-||x-x'||^2/2\sigma^2)}$. Hence the semiparametric quantile regression takes the form
\begin{equation}\label{model2}
Y_i=f_{\tau}(X_i)+\varepsilon_i,\quad f_{\tau}(x)=\sum_{j=1}^n\alpha_jk(X_j,\,x)+\beta_{\tau} x+c.
\end{equation}

For model (\ref{model2}), one important consideration is the smoothness of the curve.
More specifically, we tend to prefer smooth curves than rough ones.
Usually, the integrated squared second derivative is used to quantify the smoothness of a curve, see \cite{Green1994}.
Therefore, from \cite{Koenker2005}, to estimate $f_{\tau}(x)$, we consider the following quantile regression minimisation problem
\begin{eqnarray}\label{min1}
&&\min\left(\sum_{i=1}^n l_{\tau}\left(Y_i-f_{\tau}(X_i)\right)+\lambda\int(f_{\tau}''(x))^2\,\textrm{d}x\right)\\
&=&\min\left(\sum_{i=1}^n l_{\tau}\Big(Y_i-c_{\tau}-\sum_{j=1}^n\alpha_jk(X_j,\,X_i)-\beta_{\tau} X_i\Big)+\lambda\int\Big(\sum_{i=1}^n\alpha_ik''(X_i,\,x)\Big)^2\,\textrm{d}x\right),\nonumber
\end{eqnarray}
where $l_{\tau}$ is the loss function
\begin{eqnarray*}
l_{\tau}=\left\{
\begin{array}{cc}
\tau x,&\quad x\geq0,\\
(\tau-1)x,&\quad x<0.
\end{array}
\right.
\end{eqnarray*}
and $\lambda$ is the smoothing parameter. Denote $K=(k_{ij})$, $k_{ij}=k(X_i,\,X_j)$. Using the Riesz representation theorem,
\begin{equation*}
\int\Big(\sum_{i=1}^n\alpha_ik''(X_i,\,x)\Big)^2\,\textrm{d}x=\sum_{i,\,j}\alpha_i\,\alpha_j\,k(X_i,\,X_j)=\vec{\alpha}_{\tau}^T\,K\,\vec{\alpha}_{\tau},
\end{equation*}
where $\vec{\alpha}_{\tau}=(\alpha_1,\,\alpha_2,\,\cdots,\,\alpha_n)^T$. Hence (\ref{min1}) can be rewritten as
\begin{eqnarray}\label{min2}
\min\left\{\sum_{i=1}^n l_{\tau}\Big(Y_i-c_{\tau}-\vec{\alpha}_{\tau}^T\,K-\beta_{\tau} X_i\Big)+\lambda\,\vec{\alpha}_{\tau}^T\,K\,\vec{\alpha}_{\tau}\right\}.
\end{eqnarray}

Our main object of the current paper is on the inference for $\tau_1=97.5\%$ and $\tau_2=2.5\%$ conditional quantile curves simultaneously.
Therefore, we will focus on the minimisation problem
\begin{eqnarray}\label{min3}
\min\Big\{\sum_{i=1}^n l_{\tau_1}\big(Y_i-c_{\tau_1}-\vec{\alpha}_{\tau_1}^T\,K-\beta_{\tau_1} X_i\big)
                      +l_{\tau_2}\big(Y_i-c_{\tau_2}-\vec{\alpha}_{\tau_2}^T\,K-\beta_{\tau_2} X_i\big) \nonumber \\
                      +\lambda\,\vec{\alpha}_{\tau_1}^T\,K\,\vec{\alpha}_{\tau_1}+\lambda\,\vec{\alpha}_{\tau_1}^T\,K\,\vec{\alpha}_{\tau_1}\Big\}.
\end{eqnarray}


It can be easily shown that the minimization of the loss function $l_{\tau}$ is exactly equivalent to the maximization
of a likelihood function formed by combining independently distributed asymmetric Laplace densities ALD$(\mu,\,\tau)$,
\begin{eqnarray}\label{Laplace}
g(y|\mu,\,\tau)=\tau(1-\tau)\exp\{-l_{\tau}(y-\mu)\}.
\end{eqnarray}
Therefore, in the following, we adopt a  Baysesian approach to solve the problem (\ref{min3}).

As our approach is Bayesian, we begin by defining the prior density for parameters.
Our prior for $\vec{\alpha}$ is defined the multivariate normal density
\begin{eqnarray*}
\pi(\vec{\alpha}_{\tau}|\lambda)=\textrm{MVN}(\textbf{0},\,\lambda^{-1}K^{-}),
\end{eqnarray*}
where $K^{-}$ is the generalized inverse matrix of $K$.
We next require a priority on the smoothing parameter $\lambda$, which is constrained by a lower limit of zero.
Here, we follow \cite{Thompson2010} by using the gamma density as our prior for $\lambda$
\begin{eqnarray*}
\pi(\lambda)=\frac{\lambda^{\xi-1}\exp(-\lambda/\theta)}{\Gamma(\xi)\,\theta^{\xi}},
\end{eqnarray*}
where $\xi,\,\theta$ are hyperparameters. Under this prior $\textrm{E}\lambda=\xi\theta$,
and $\textrm{Var}(\lambda)=\xi\theta^2$, results can be used to guide hyperparameter choice.

Assume $Y_i$ follows asymmetric Laplace distribution (\ref{Laplace}) ALD $(f_{\tau_k}(X_i),\,\tau_k)$, $k=1,\,2$, the likelihood for $n$ independent observations is given by
\begin{eqnarray}\label{likelihood}
&&L(Y|\vec{\alpha}_{\tau_1},\,\vec{\alpha}_{\tau_2},\,\beta_{\tau_2},\,\beta_{\tau_1},\,c_{\tau_1},\,c_{\tau_2})\\
&=&\tau_1^n(1-\tau_1)^n \tau_2^n(1-\tau_2)^n \exp\left\{-\sum_{i=1}^n l_{\tau_1}(Y_i-f_{\tau_1}(X_i))-\sum_{i=1}^n l_{\tau_2}(Y_i-f_{\tau_2}(X_i))\right\}\nonumber
\end{eqnarray}
Using uniform distribution as the improper prior density for $\beta_{\tau_1},\beta_{\tau_2},\,c_{\tau_1},\,c_{\tau_2}$,
combining $\pi(\lambda)$, $\pi(\vec{\alpha}_{\tau_1}|\lambda)$, $\pi(\vec{\alpha}_{\tau_2}|\lambda)$,
and $L(Y|\vec{\alpha}_{\tau_1},\,\vec{\alpha}_{\tau_2},\,\beta_{\tau_2},\,\beta_{\tau_1},\,c_{\tau_1},\,c_{\tau_2})$,
we can write the posterior density function of parameters $\vec{\alpha}_{\tau_1},\,\vec{\alpha}_{\tau_2},\,\beta_{\tau_2},\,\beta_{\tau_1},\,c_{\tau_1},\,c_{\tau_2},\,\lambda$
\begin{eqnarray}\label{post}
&&\pi(\vec{\alpha}_{\tau_1},\,\vec{\alpha}_{\tau_2},\,\beta_{\tau_2},\,\beta_{\tau_1},\,c_{\tau_1},\,c_{\tau_2},\,\lambda|Y)\nonumber\\
&\propto& L(Y|\vec{\alpha}_{\tau_1},\,\vec{\alpha}_{\tau_2},\,\beta_{\tau_2},\,\beta_{\tau_1},\,c_{\tau_1},\,c_{\tau_2})\,\pi(\vec{\alpha}_{\tau_1}|\lambda)\,\pi(\vec{\alpha}_{\tau_2}|\lambda)\,\pi(\lambda)
  \,\pi(\beta_{\tau_1})\,\pi(\beta_{\tau_2})\,\pi(c_{\tau_1})\,\pi(c_{\tau_2},)\nonumber\\
&\propto&\exp\Big\{-\sum_{i=1}^n l_{\tau_1}\Big(Y_i-c_{\tau_1}-\vec{\alpha}_{\tau_1}^T\,K-\beta_{\tau_1} X_i\Big)-\sum_{i=1}^n l_{\tau_2}\Big(Y_i-c_{\tau_2}-\vec{\alpha}_{\tau_2}^T\,K-\beta_{\tau_2} X_i\Big)\Big\}\nonumber\\
&&\exp\Big\{-\lambda\,\vec{\alpha}_{\tau_1}^T\,K\,\vec{\alpha}_{\tau_1}-\lambda\,\vec{\alpha}_{\tau_2}^T\,K\,\vec{\alpha}_{\tau_2}\Big\}\pi(\lambda)
\end{eqnarray}


Next we shall use Metropolis-Hastings algorithm and Gibbs Sampler to obtain Markov Chain Monte Carlo (MCMC) samples from the posterior distribution (\ref{post}). \cite{Brooks1998} introduced a multivariate statistics, which is called Multivariate Potential Scale Reduction Factor (MPSRF),
that can be applied here to test the convergence of MCMC chains. We set the number of iterations $d$ to $2L$ and promise a burn-in of $L$ iterations. Inference is based on thinned values of $\vec{\alpha}_{\tau_1},\,\vec{\alpha}_{\tau_2},\,\beta_{\tau_2},\,\beta_{\tau_1},\,c_{\tau_1},\,c_{\tau_2}$
after burn-in.

\subsection{Hypothesis Test}

We consider the null hypothesis $H_0$: the ADS of mutant type is normal, against the alternative hypothesis $H_1$: the ADS of mutant type is abnormal. Firstly, we use Bayesian semiparametric quantile regression to get the estimators of quantile curves $\hat{f}_{0.025}(x)$ and $\hat{f}_{0.975}(x)$, which are the lower and upper confidence bounds for wild-type embryos respectively. Within these curves, we get a $95\%$ confidence interval. If the ADS of mutant individual is located in the interval at corrrsponding life time of mother cell, we accept $H_0$. If not, we reject $H_0$.


\section{Application}\label{application}

\subsection{Synthesized Data}

Here we report a simulation study designed to evaluate the performance of the Bayesian quantile regression model.
Set $\tau_1=97.5\%$ and $\tau_2=2.5\%$. Different quantile curves are considered in this section.

\begin{tabular}{lll}
Scenario (1):  & $f_1(x)=a_1\,x+5,$        & $f_2(x)=-1,$   \\
               & $ a_1\sim U(0.1,\,0.5).$  &                \\
Scenario (2):  & $f_1(x)=a_1\,x+5,$        & $f_2(x)=a_2\,x-1,$\\
               & $a_1\sim U(0.1,\,0.5),$   & $a_2\sim U(-0.5,\,-0.1)$.\\
Scenario (3):  & $f_1(x)=a_1\,x+5,$        & $f_2(x)=a_2\,x-1,$\\
               & $a_1\sim U(0.1,\,0.5),$   & $a_2\sim U(0.1,\,0.2).$\\
Scenario (4):  & $f_1(x)=a_1\,x^2+5,$      & $f_2(x)=-1,$\\
               & $a_1\sim U(0.01,\,0.02).$ &                    \\
Scenario (5):  & $f_1(x)=a_1\,x^2+5,$      & $f_2(x)=a_2\,x-1,$ \\
               & $a_1\sim U(0.01,\,0.02),$ & $a_2\sim U(0.1,\,0.2).$\\
Scenario (6):  & $f_1(x)=5,$               & $f_2(x)=-5$.\\\\
\end{tabular}

For each scenario, we generate the data as follows.
Firstly, we select 100 different $\mathcal{D}_{i_1},\,\mathcal{D}_{i_2},\,\cdots,\,\mathcal{D}_{i_{100}}$ from 351 datasets $\mathcal{D}_1,\,\mathcal{D}_2,\,\cdots,\,\mathcal{D}_{351}$.
Secondly, for each dataset $\mathcal{D}_{k}=\{(X_{k1},\,Y_{k1}),\,(X_{k2},\,Y_{k2}),\,\cdots,\,(X_{kn_k},\,Y_{kn_k})\}$,
delete the data $Y$, and keep $X$ as the synthesized one. Then given $X_i,\,(i=1,\,2,\,\cdots,\,n)$, we simulate $Y_{i}$ from a Normal distribution N$(\mu_i,\,\sigma^2_i)$, where
$$\mu_i=\frac{f_1(X_i)+f_2(X_i)}{2},\qquad \sigma_i=\frac{f_1(X_i)-f_2(X_i)}{2\Phi^{-1}(\tau_1)}$$
$\Phi(\cdot)$ is the cumulative distribution function of the standard normal distribution.
This setting guarantees that the $\tau_1$ and $\tau_2$ conditional quantile curves of $Y_i$ are $f_1(X_i)$ and $f_2(X_i)$.
For each dataset $\mathcal{D}_{k}$, repeat the above process 10 times. Hence we get 1000 synthesized datasets.

\begin{figure}[htbp]
\includegraphics[scale=1]{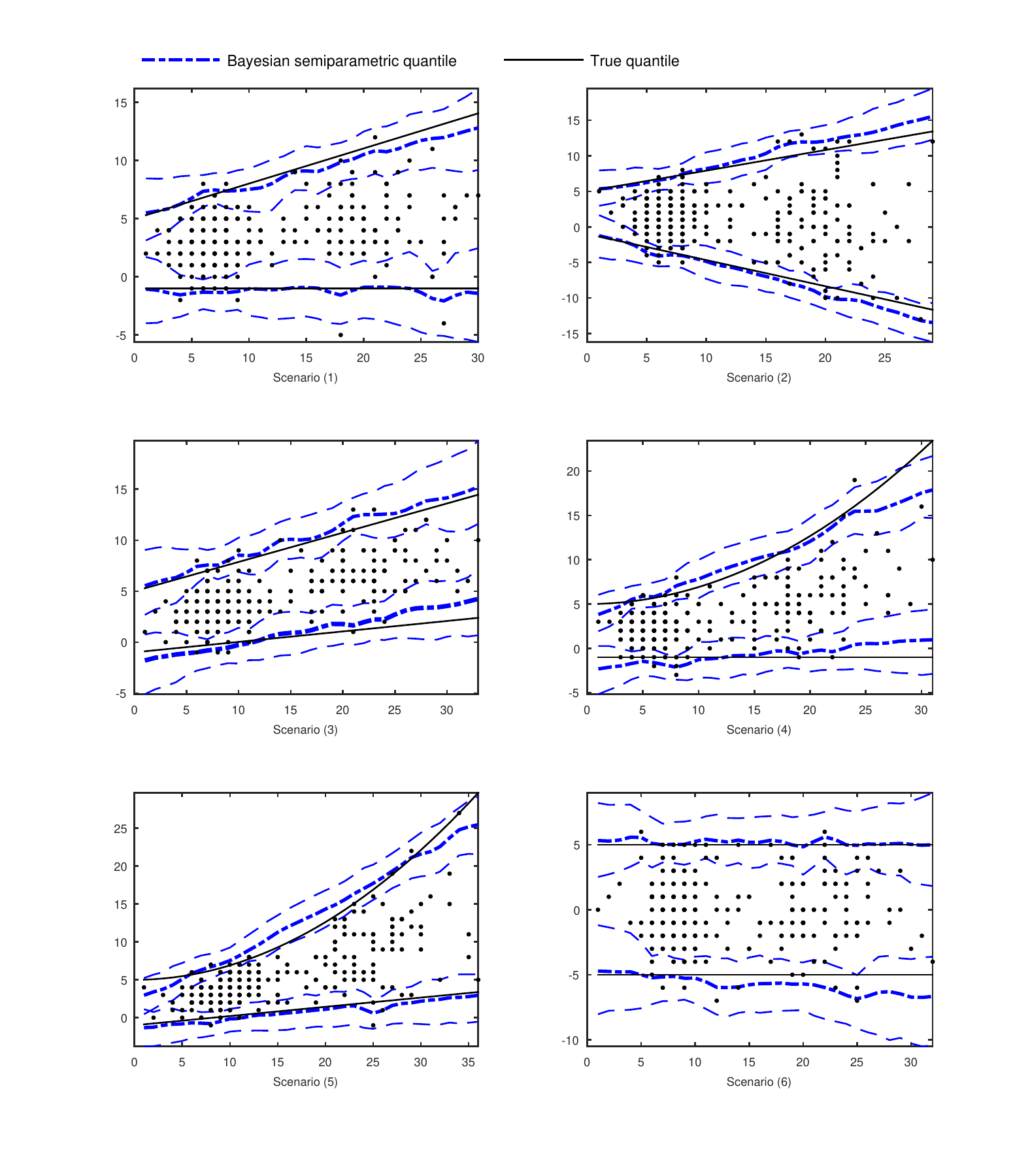}
\caption{Fitting examples of Scenarios 1-6. Black solid line represents true quantile and blue bold line indicates mean estimation results of our semiparametric method with thinned blue lines standing for $95\%$ confidence inetrval for each estimated quantile curve.}\label{simu}
\end{figure}

Each time, we run two independent MCMC chains and then evaluate the MPSRF to monitor the convergence of Markov chains.
After the Markov chains converge, we get the mean estimator of $\hat{f}_{\tau_1}(x)$ and $\hat{f}_{\tau_2}(x)$ at all $X_i$ points. Table~\ref{scenario} shows the coverage rates (CR) of $95\%$ confidence intervals of $f_{\tau_1}(x)$ and $f_{\tau_2}(x)$. We list the averaged Mean Square Error (A MSE) of $\hat{f}_{\tau_1}(x)$ and $\hat{f}_{\tau_2}(x)$ and their variances (V MSE)over all synthesized data sets within each scenario in Table~\ref{scenario}. \cite{Ho2015} discards the possible correlation between $X$ and $Y$ and directly use $2.5\%$ and $97.5\%$ quantiles of all points $Y$ to define the $95\%$ confidence interval and the results are also listed in Table~\ref{scenario} for comparison.
\footnotesize
\begin{table}
\centerline{
\begin{tabular}{|cc|c|cc|cc|}
\multicolumn{7}{c}{Table 2.}\\\hline
                &                      &             &  \multicolumn{2}{c|}{Our Method}&\multicolumn{2}{c|}{\cite{Ho2015}}\\\cline{3-7}
                &                      &CR&   A(MSE)  &     V(MSE)       &A(MSE)&   V(MSE)          \\\hline
 Scenario 1     &  $f_{\tau_1}(x)$     & $99.48\%$   &    0.5999       &         0.6234       &   8.1659    &  50.2678  \\
                &  $f_{\tau_2}(x)$     & $99.29\%$   &    0.5999       &         0.4862       &   0.1917    &  0.1393    \\\hline
 Scenario 2     &  $f_{\tau_1}(x)$     & $97.36\%$   &    1.2797       &         3.4002       &   9.1810    &  55.6763  \\
                &  $f_{\tau_2}(x)$     & $97.79\%$   &    1.1438       &         2.1626       &   9.0735    &  54.2697  \\\hline
 Scenario 3     &  $f_{\tau_1}(x)$     & $99.65\%$   &    0.4768       &         0.2063       &   9.4806    &  63.3023 \\
                &  $f_{\tau_2}(x)$     & $99.67\%$   &    0.4684       &         0.1870       &   3.5869    &  5.6692 \\\hline
 Scenario 4     &  $f_{\tau_1}(x)$     & $90.81\%$   &    2.3845       &         5.1563       &   27.9158   &  651.5690\\
                &  $f_{\tau_2}(x)$     & $98.95\%$   &    0.7315       &         0.8093       &   0.1931    &  0.1570\\\hline
 Scenario 5     &  $f_{\tau_1}(x)$     & $92.37\%$   &    1.9928       &         3.6880       &   28.8412   &  671.9742\\
                &  $f_{\tau_2}(x)$     & $99.56\%$   &    0.5295       &         0.3305       &   3.7085    &  5.3046\\\hline
 Scenario 6     &  $f_{\tau_1}(x)$     & $99.38\%$   &    0.5264       &         0.2746       &   0.2869    &  0.1938\\
                &  $f_{\tau_2}(x)$     & $99.60\%$   &    0.4916       &         0.2298       &   0.2405    &  0.1988\\\hline
\end{tabular}}
\caption{Estimation results of our semiparametric method and \cite{Ho2015}}.\label{scenario}
\end{table}\normalsize
As expected, our method is much more accurate than \cite{Ho2015} in most scenarios except when $f_{\tau}(x)$ is a constant function which meets \cite{Ho2015} perfectly. Figure~\ref{simu} demonstrates a fitting example of each scenario.

\subsection{Real Data}

In this section, we apply our method to all the ADS of 84 different mutant-type individuals. We also apply method from \cite{Ho2015} to do the comparison of detected results. We list part of the results in Table~\ref{realresult}, and complete results can be found in Supplementary Materials.
\footnotesize
\begin{table}
\begin{center}
\begin{longtable}{c|cccccccccccccccc|}\hline
  & \rotatebox{270}{ABalpaap} & \rotatebox{270}{ABarppaa} & \rotatebox{270}{ABarpppa} & \rotatebox{270}{ABarpaa} & \rotatebox{270}{ABplpapp} & \rotatebox{270}{ABarpapp} & \rotatebox{270}{ABplaaap} & \rotatebox{270}{ABalapap} & \rotatebox{270}{ABalaap} & \rotatebox{270}{ABalaapp} & \rotatebox{270}{ABalappa} & \rotatebox{270}{ABalapp} & \rotatebox{270}{ABalppa} & \rotatebox{270}{ABaraaaa} & \rotatebox{270}{ABarapaa}\\\hline
\textit{pop1}&	  $\surd$&	&	&	$\diamond$&	&	&	&	&	&	&	&	&	&	$\bullet$&	  \\ \hline
\textit{gei17}&	  &	&	&	$\surd$&	&	&	&	&	&	&	$\surd$&	&	$\surd$&	&	  \\ \hline
\textit{egl27}&	  &	   &	&	&	&	$\surd$&	&	&	&	&	&	&	&	$\surd$&	  \\ \hline
\textit{ceh43}&	  &	   &	&	&	&	&	&	&	$\surd$&	$\diamond$&	&	$\surd$&	&	$\diamond$&	  \\ \hline
\textit{tbp1}&	  &	&	&	&	$\surd$&	&	$\surd$&	&	&	$\surd$&	&	$\diamond$&	$\diamond$&	&	  \\ \hline
\textit{tbx37}+\textit{tbx38}& &	&	&	&	&	$\diamond$&	&	&	&	&	&	$\diamond$&	&	&	  \\ \hline
\textit{cdk8}&	  &	&	&	&	&	$\surd$&	&	$\surd$&	&	&	&	&	&	&	  \\ \hline
\textit{F11A10-1}& &	 &	&	$\surd$&	&	$\surd$&	&	$\surd$&	&	&	$\surd$&	&	&	$\diamond$&	  \\ \hline
\textit{lin12}+\textit{glp1}&	$\diamond$&	&	&	&	&	&	&	&	&	&	&	&	$\diamond$&	$\bullet$&	  \\ \hline
\textit{lin23}&	  &	&	&	&	&	$\surd$&	&	$\surd$&	&	$\surd$&	$\surd$&	&	&	&	  \\ \hline
\textit{ceh13}&	   &	$\diamond$&	&	$\surd$&	&	&	$\surd$&	&	&	&	&	&   &	&	$\surd$  \\ \hline
\textit{epc1}&	  $\diamond$&	&	&	&	&	&	&	&	&	$\diamond$&	$\surd$&	$\surd$&	&	$\diamond$&	  \\ \hline
\textit{frg1}&	   &	&	&	&	&	&	$\surd$&	&	&	&	$\surd$&	&	&	&	  \\ \hline
\textit{snfc5}&	  $\diamond$&	$\diamond$&	$\diamond$&	$\diamond$&	$\diamond$&	$\diamond$&	$\surd$&	$\diamond$&	$\surd$&	&	&	$\diamond$&	&	$\bullet$&	$\diamond$  \\ \hline
\textit{dpy28}&	   &	 &	&	&	&	&	$\surd$&	&	&	$\diamond$&	$\surd$&	&	&	$\diamond$&	$\surd$  \\ \hline
\textit{cbp1}&	  $\diamond$&	$\diamond$&	$\diamond$&	$\diamond$&	&	&	$\diamond$&	$\diamond$&	$\diamond$&	$\bullet$&	&	$\diamond$&	$\surd$&	$\surd$&	$\diamond$  \\ \hline
\textit{let526}&   $\surd$&	$\diamond$&	&	&	&	&	$\diamond$&	$\diamond$&	&	&	$\diamond$&	&	&	&	$\surd$  \\ \hline
\textit{cblin40}&   &	&	&	$\surd$&	&	&	&	&	&	&	&	&	&	$\bullet$&	  \\ \hline
\caption{Comparsion of detected results between our method and \cite{Ho2015}. First column represents mutant types and first row represents different types of ADSs. $\surd$ indicates abnormal one detected only by our method while $\bullet$ indicating significance only detected by \cite{Ho2015} with $\diamond$ represnting significant ADS detected by both methods.}\label{realresult}
\end{longtable}
\end{center}
\end{table}
\normalsize

\begin{figure}[htbp]
\centering
\includegraphics[scale=0.5]{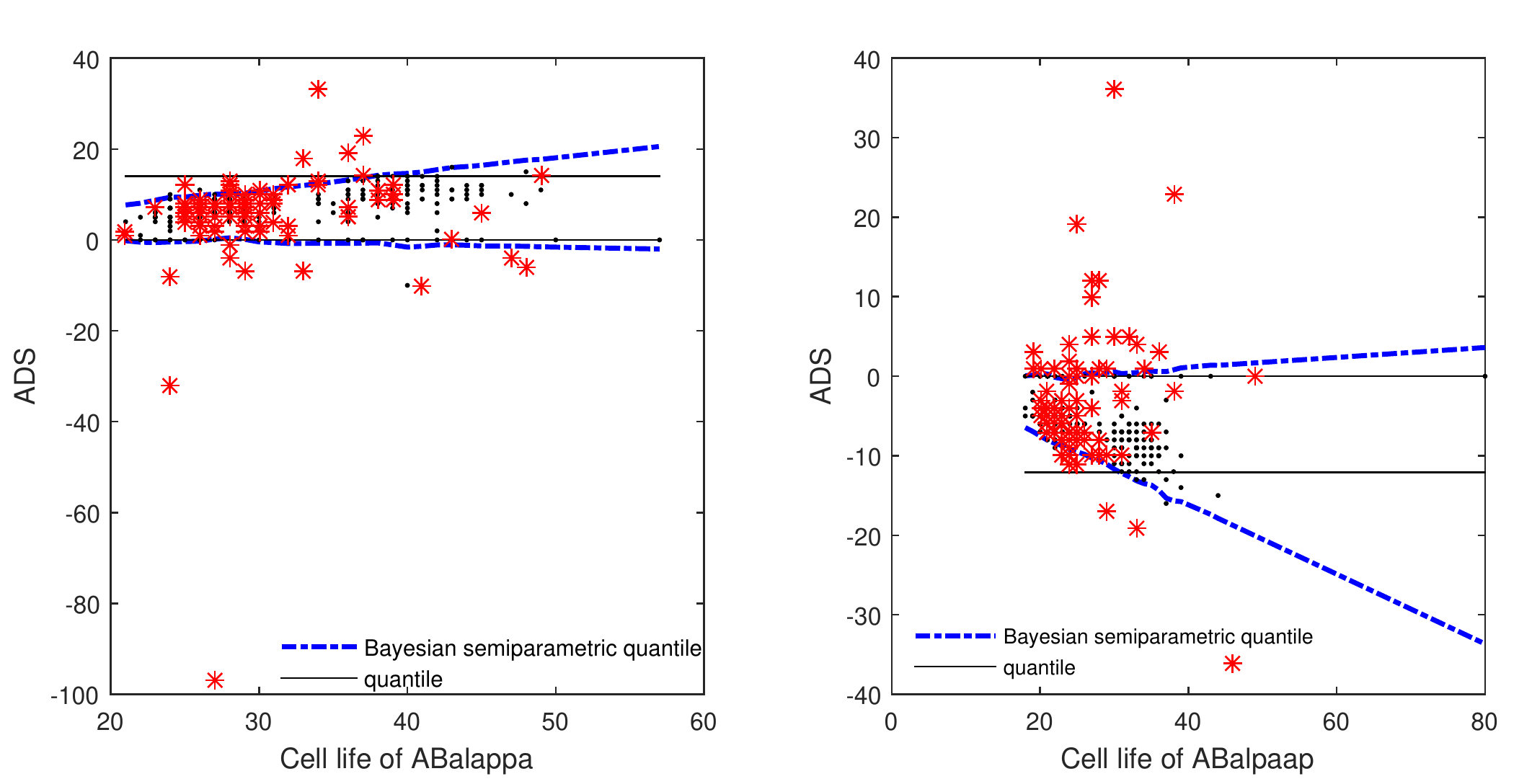}
\caption{Black dots are the wild-type sample, and the red stars represent the mutant-type individuals.}\label{RealData} 
\end{figure}

Figure~\ref{RealData} is a typical example, showing the difference between two methods.

Furthermore, we binarize the results with $1$ indicating abnormal ADS detected by our method and summarize them into a binary vector for each mutant type with each entry corresponding to one type of ADS. Then we use the data to cluster the mutant genes into 3 types by kmodes method proposed by \cite{Huang1997} for discrete data. We analyze the significance of our clustering result by Gene Set Enrichment Analysis through DAVID developed by \cite{Huang2009a,Huang2009b}. Table~\ref{go} demonstrates the analysis results and the details of clustering result can be found in Appendix~\ref{cluster}. Type 3 is not significant in any functional module which may be derived from only 4 genes involved. 

\begin{table}
	\centerline{
		\begin{tabular}{|c|c|c|c|}
			\hline
			Type&Category&Term&P-value\\
			\hline
			1&BP&Reproduction&$0.028$\\
			\hline
			\multirow{3}{*}{2}&BP&Body morphogenesis&$0.030$\\
			\cline{2-4}
			&BP&DNA binding&$0.018$\\
			\cline{2-4}
			&BP&Transcription factor activity, sequence-specific DNA binding&$0.039$\\
			\hline
		\end{tabular}
	}
	\caption{Gene Set Enrichment Analysis. Significant P-values smaller than $0.05$ are demonstrated in the table.}\label{go}
\end{table}

\section{Conclusion}\label{conclusion}
We provide a principled automatic procedure to detect abnormal ADS of \textit{C. elegans} mutant individuals. Simulation studies and real data examples show that our method can estimate the quantile curves precisely and detect the significant outliers efficiently. Gene Enrichment Analysis shows that our clustering results based on abnormal ADS make sense to a certain extent which also validates the importance of ADS. In general, our method can handle most cases well except for the case where the real quantile curve violates our model assumption severely, such as exponential or high-order polynomial function. A non-parametric model may be a good choice for this situation. However, the unsmooth problem of fitted curve is a considerable challenge.

\section{Appendix}

\subsection{Spearman's rho}\label{a1}

Spearman's rank correlation coefficient or Spearman's rho, named after Charles Spearman, often denoted by $\rho$ or as $r_s$,
is a nonparametric measure of rank correlation. It assesses how well the relationship between two variables can be described using a monotonic function.
Suppose $(X_1,\,Y_1),\,(X_2,\,Y_2),\,\cdots,\,(X_n,\,Y_n)$ is an i.i.d. copy of $(X,\,Y)$, and convert $(X_i,\,Y_i)$ to ranks $(R_i,\,S_i)$,
then Spearman's rho is computed from
$$\rho=1-\frac{6\sum_{i=1}^nd_i^2}{n(n^2-1)}$$
where
$d_i=R_i-S_i$ is the difference between rankings.

When ties are present in rankings, we should use average ranking instead of ranking.
Let $u_1,\,u_2,\,\cdots$ and $v_1,\,v_2,\,\cdots$ be the numbers of each ties in $X$ and $Y$ respectively.
Denote $U=\sum(u_j^3-u_j)$, $V=\sum(v_j^3-v_j)$. The adjusted Spearman's rho is showed as follow,
$$\rho=\frac{n(n^2-1)-6\sum_{i=1}^nd_i^2-(U+V)/2}{\sqrt{n(n^2-1)-U}\sqrt{n(n^2-1)-V}}.$$

The sign of the Spearman correlation indicates the direction of association between $X$ and $Y$.
If $Y$ tends to increase when $X$ increases, the Spearman correlation coefficient is positive. If $Y$ tends to decrease when $X$ increases, the Spearman correlation coefficient is negative.
A Spearman correlation of zero indicates that there is no tendency for $Y$ to either increase or decrease when $X$ increases.
When $X$ and $Y$ are perfectly monotonically related, the Spearman correlation coefficient becomes 1.

\subsection{Pearson correlation and Spearman's correlation of synthesized Data}\label{cor}
\begin{figure}[htbp]
\centering
\includegraphics[scale=1]{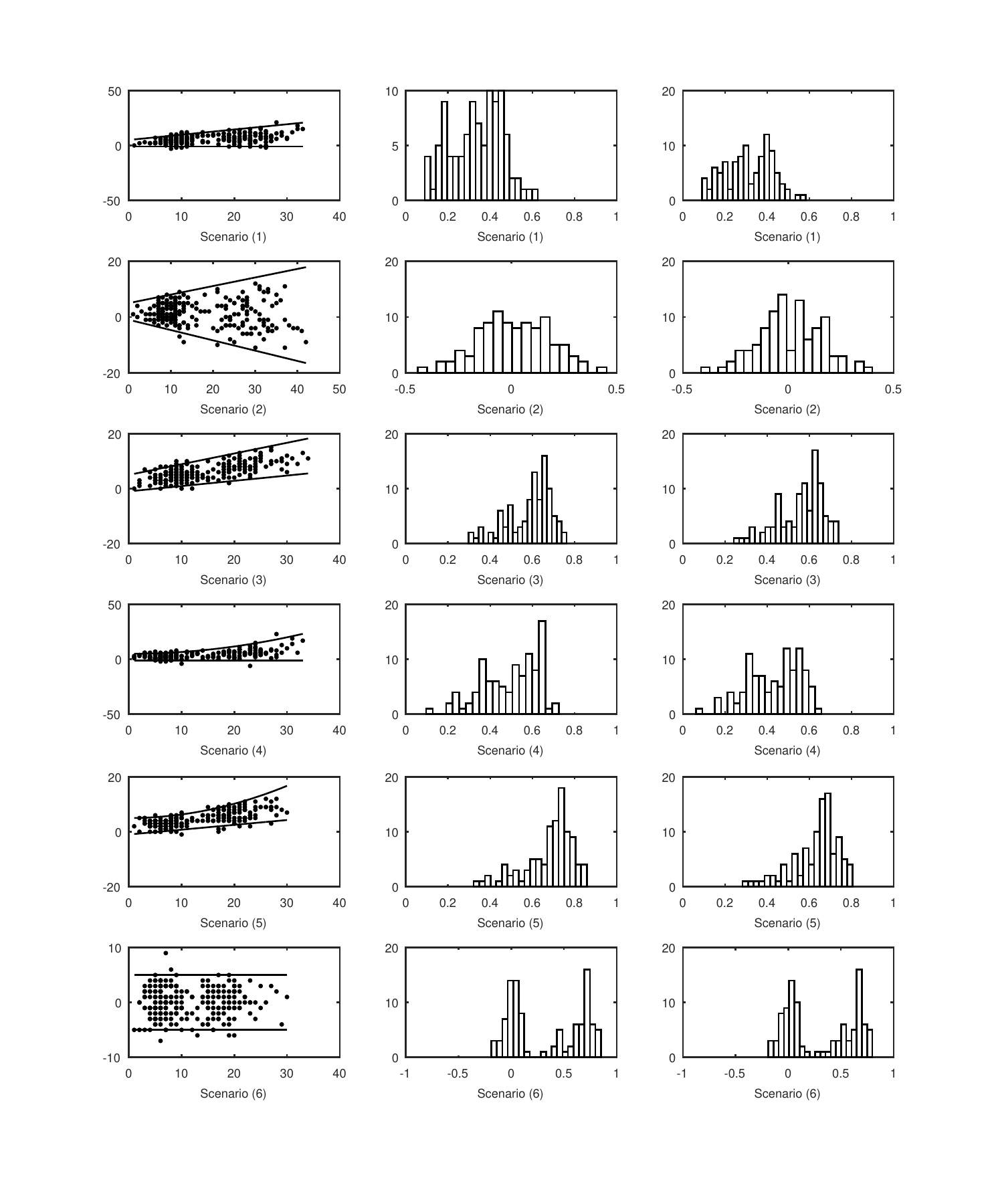}
\caption{Scenario 1--6: Left column is an example of synthesized data set under each scenario. Middle and right columns show corresponding Pearson correlation and Spearman's correlation respectively}\label{correlation}
\end{figure}

For each scenario, we calculate the Pearson correlation and Spearman's $\rho$ for synthesized datasets. Figure~\ref{correlation} shows the histograms of Pearson correlation and Spearman's correlation under different Scenarios.

\subsection{MCMC convergence}

Visual assessment of the convergence of the Metropolis-Hastings algorithm is shown as following.
\begin{figure}[htbp]
\begin{center}
\includegraphics[scale=0.75]{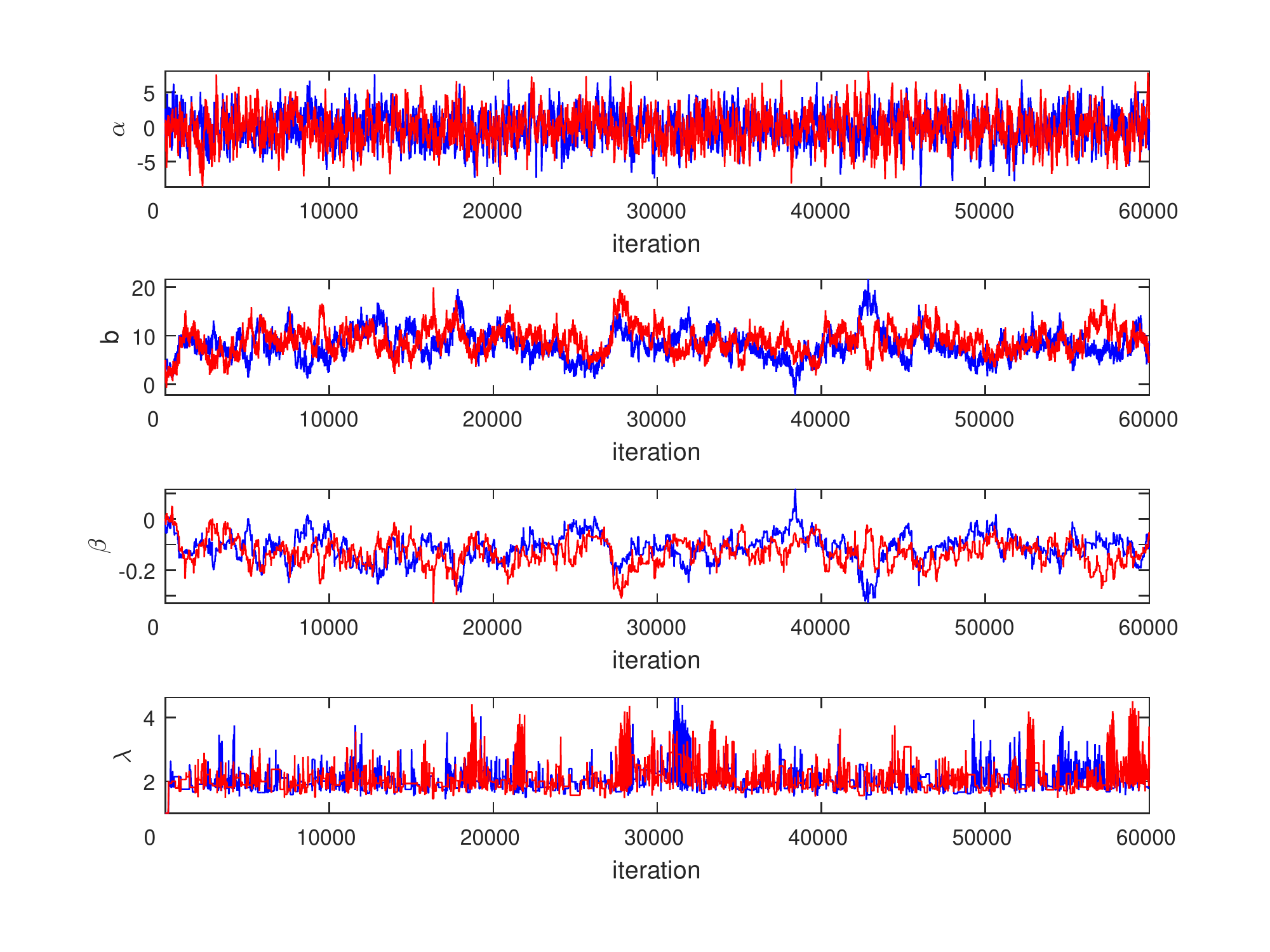}
\end{center}
\caption{Plot of iteration traces of $\alpha$, b, $\beta$, $\lambda$.}\label{Chain}
\end{figure}

Figure~\ref{Chain} shows that the time series plots of $\alpha$, b, $\beta$, $\lambda$ iteration traces.
After initially examining the plots of the MCMC chain, we used
the more formal MPSRF statistic to assess convergence of the MCMC chain. This statistic compares the variances between and within
chains to monitor convergence. The value of MPSRF should certainly not exceed 1.2 as suggested in the literature.
In fact, after 80000 interation, we get that MPSRF take values nearly 1.101.
The time series plots together with satisfactory values of the MPSRF statistic gave us confidence
that the Metropolis-Hastings algorithm was producing realizations from the posterior distribution
$\pi(\vec{\alpha}_{\tau_1},\,\vec{\alpha}_{\tau_2},\,\beta_{\tau_1},\,\beta_{\tau_2},\,c_{\tau_1},\,c_{\tau_2},\,\lambda|Y)$.

\subsection{Gene Clustering}\label{cluster}
Details of clustering are demonstrated in Table~\ref{cluster}.
\begin{table}[!h]
	\centerline{
		\begin{tabular}{|c|c|c|}
			\hline
			Type&Count&Genes\\
			\hline
			1&12&\textit{gad-1,Y55F3AM.3,cul-1,lit-1,ykt-6,hlh-2,dsh-2,ubc-9,cbp-1,ddx-23,gsk-3,wrm-1}\\
			\hline
			\multirow{4}{*}{2}&\multirow{4}{*}{49}&\textit{pie-1,cogc-2,sptf-3,abce-1,pal-1,pop-1,mex-6,unc-62,grh-1,lag-1,nhr-25,frg-1,epc-1,}\\
			&&\textit{egl-18,gei-17,ref-1,tbx-33,egl-27,efl-1,die-1,F57C9.4,ceh-43,rps-9,W06E11.1,tbp-1,}\\
			&&\textit{dpy-22,vps-37,cdk-8,F13H8.9,kin-19,npp-2,src-1,pes-1,mom-2,lin-23,pfd-1,ceh-13,}\\
			&&\textit{mom-5,M03C11.1,mdt-11,plp-1,dpy-28,mys-2,par-6,let-526,ham-1,skr-2,arx-1,cpt-2,}\\
			\hline
			3&4&\textit{mex-1,par-2,sel-8,snfc-5}\\
			\hline
		\end{tabular}
	}
	\caption{Clustering results of 65 genes.}\label{cluster}
\end{table}

\clearpage \addcontentsline{toc}{section}{Bibliography}
\nocite{}


\end{document}